\documentclass{PoS}
\usepackage{graphics}
\usepackage{natbib}
\usepackage{aastex_hack}

\title{NuGrid: Toward High Precision Double-Degenerate Merger Simulations with SPH in 3D}

\ShortTitle{NuGrid: DD Mergers with SPH}

\author{
\speaker{Steven Diehl}$^{abc}$,
Christopher L. Fryer$^{ac}$, 
Aimee Hungerford$^{ac}$,
Gabriel Rockefeller$^{ac}$,
Michael Bennett$^{ad}$,
Falk Herwig$^{ade}$,
Raphael Hirschi$^{ad}$,
Marco Pignatari$^{adf}$,
Georgios Magkotsios$^{afg}$,
Francis X. Timmes$^{ag}$,
Patrick Young$^{ag}$,
Geoffrey C. Clayton$^h$,
Patrick Motl$^hi$ 
and
Joel E. Tohline$^h$\\ 
\llap{$^a$}The NuGrid Collaboration\\
\llap{$^b$}Theoretical Astrophysics Group (T-6), Los Alamos National Laboratory, Los Alamos, NM, 87544, USA\\
\llap{$^c$}Computational Methods (CCS-2), Los Alamos National Laboratory, Los Alamos, NM, 87544, USA\\
\llap{$^d$}Astrophysics Group, Keele University, ST5 5BG, UK\\
\llap{$^e$}Dept. of Physics \& Astronomy, Victoria, BC, V8W 3P6, Canada\\
\llap{$^f$} Joint Institute for Nuclear Astrophysics, University of Notre Dame, IN, 46556, USA\\
\llap{$^g$}School of Earth and Space Exploration, Arizona State University, Tempe, AZ 85287, USA\\
\llap{$^h$}Department of Physics \& Astronomy, Louisiana State University, Baton Rouge, LA 70803, USA \\
\llap{$^i$} Department of Natural, Mathematical and Information Sciences
Indiana University Kokomo, Kokomo, IN 46902, USA \\
E-mail:\email{diehl@lanl.gov}
}

\abstract{We present preliminary results from recent high-resolution double-degenerate merger simulations with the Smooth Particle Hydrodynamics (SPH) technique. We put particular emphasis on verification and validation in our effort and show the importance of details in the initial condition setup for the final outcome of the simulation. We also stress the dynamical importance of including shocks in the simulations. These results represent a first step toward a suite of simulations that will shed light on the question whether double-degenerate mergers are a viable path toward type 1a supernovae. In future simulations, we will make use of the capabilities of the NuGrid collaboration in post-processing SPH particle trajectories with a complete nuclear network to follow the detailed nuclear reactions during the dynamic merger phase.}

\FullConference{10th Symposium on Nuclei in the Cosmos\\
		 July 27 - August 1 2008\\
		 Mackinac Island, Michigan, USA}

\begin{document}

\section{Introduction}

Double-Degenerate (DD) mergers are an alternative to the single-degenerate progenitor scenario for type 1a supernovae (SN) that has recently received a surge in attention \citep[e.g.][]{Saio2004,YoonDDSPH}. In this process, two white dwarfs (WD) merge, pushing one of them over the Chandrasekhar limit, which causes it to contract and finally explode. One of the main advantages of this model is that the predicted rate of such DD merger events is consistent with the observed SN rate, in contrast to many other SN1a progenitor models. This DD merger scenario was early on ruled out by \citet{Nomoto}. They pointed out that a CO WD accreting at rates higher than $\sim10^{-6}M_{\sun}\,s^{-1}$ would undergo ``accretion induced collapse'', turning it into an Oxygen/Magnesium/Neon WD. When pushed over the Chandrasekhar limit, this WD would then be able to cool sufficiently through neutrino emission to avoid a type 1a SN, forming a neutron star instead. 

However, Nomoto \& Kondo's study assumed constant accretion rates, and did not account for a complex rotation profile. Promising new results have shed a new light on their paradigm, opening up small windows of opportunity for SN1a as a result of DD mergers \citep{YoonDDSPH}. What is now needed are realistic high-precision simulations to determine if DD mergers are indeed a viable path to SN1a. Our Goal is to conduct a rigorous verification and validation process and produce high-precision SPH simulations to answer these questions, using a modified version of the SNSPH code \citep{SNSPH,DiehlHYDEF07}. In this work, we put particular emphasis on the importance of appropriate initial conditions and the effects of shocks in the equation of state.

\section{The Importance of Initial Conditions}

We use the self-consistent field method to setup up our initial conditions. This method was first developed by \citet{HachisuDDsetup} and employed by \citet{TohlineDD1}, \citet{TohlineDD2} and \citet{TohlineDD3}. It iteratively solves the equilibrium configuration for unequal mass close binaries in co-rotation. Its biggest advantage is that initial conditions are in perfect equilibrium, imposing very few artifacts in the simulations due to an imprecise initial setup. Its main drawback at the moment is that one is bound to choose a rather simple equation of state to produce polytropes, though work is underway to generalize this method for a more realistic equation of state.

\begin{figure}
\hfil\includegraphics[width=0.75\textwidth]{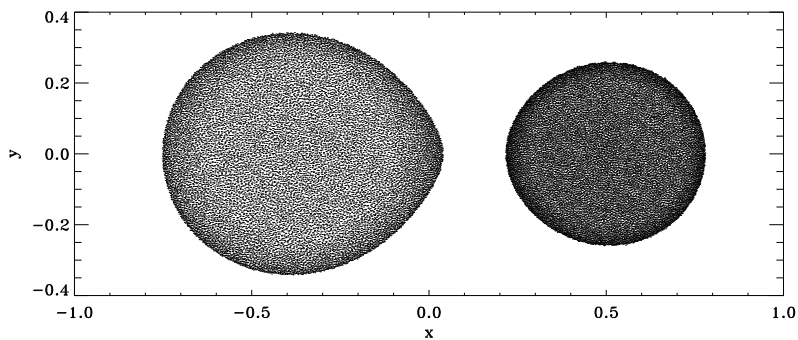}\hfil
\caption{Initial Condition Setup for a $q=1.3$ mass ratio white dwarf binary system.\label{f.wvtsetup}}
\end{figure}

In order to be able to model a rather gentle Roche lobe overflow correctly with a particle based method like SPH, we have developed a new SPH particle setup method that allows us to increase the resolution in the outer layers, while keeping an optimal distribution of particles. This new setup method%\footnote{The WVT setup method is available at ***}
is based on weighted Voronoi tesselations (WVT), allows an arbitrary spatial configuration without imposing a lattic geometry, and minimizes particle noise and density fluctuations \citep{DiehlWVTSPH}. Figure \ref{f.wvtsetup} shows an example setup for a $q=0.4$ mass ratio binary. The accretor is set up with a constant particle density on the left, while the donor on the right (filling 99.7\% of its Roche lobe) has an 8x higher resolution in its outer layers. 

We also note a strong difference in behavior for corotating vs. non-corotating systems. Non-corotating systems tend to transfer a significant fraction of the orbital angular momentum into spinning up stars. This results in the orbit shrinking, bringing the star into deeper Roche lobe contact and artificially shortening the merger time scale. For more details, refer to \citet{FryerHYDEF07}.

\section{The Importance of the Equation of State: Shocks vs. No Shocks}

\begin{figure}
\begin{center}
\includegraphics[width=0.4\textwidth]{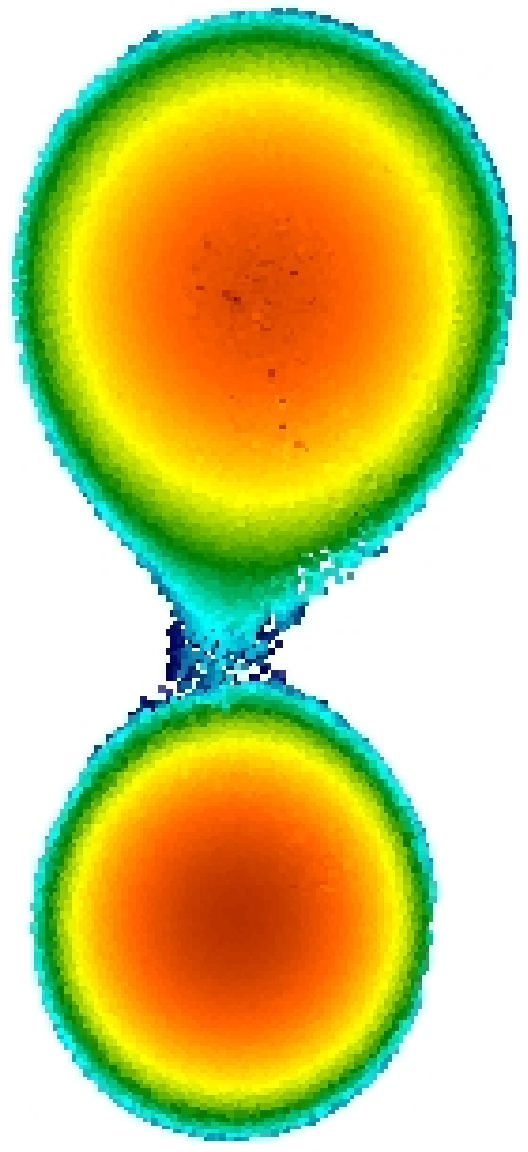}
\includegraphics[width=0.4\textwidth]{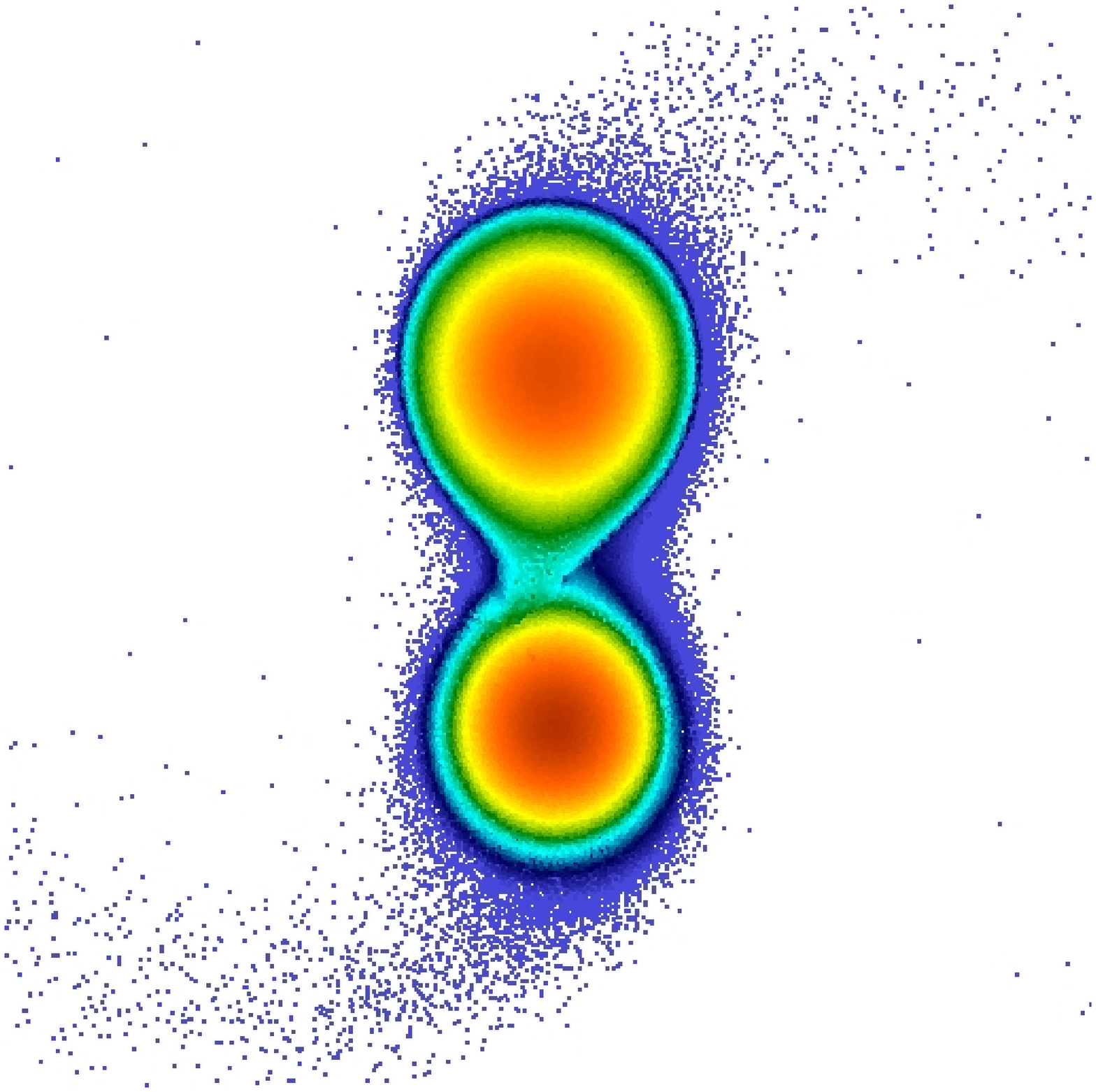}
\end{center}
\caption{Comparison of the same time step of two otherwise identical runs for an initial mass ration of $q=1.3$, only differing in their equations of state. SPH particles are colored according to their density, ranging logarithmically from $10^{-4}$ to $10^{0}$. The left side uses the the polytropic equation of state, which keeps entropy constant during the accretion process. The right side uses an ideal gas equation of state to include the effects of shocks. The donor (top) material now gets heavily shocked as it hits the accretor (bottom), which leads to the buildup of a halo that engulfes both stars. Due to our new setup method, we resolve the accretion stream with well over a thousand SPH particles on average. Also note that the binary now loses significant amounts of mass on the backside of both white dwarfs. \label{f.eos}}
\end{figure}

We have conducted various runs with different equations of state but otherwise identical initial conditions to determine the effect of the equation of state on the dynamics of the interaction. Figure \ref{f.eos} shows snapshots of two simulations with different equations of state at the same time in the simulation. The left side shows a polytropic equation of state, such that the pressure is a simple function of the density $P=k\,\rho^\Gamma$. During the simulations, this essentially results in the absence of shocks, since the entropy of the gas is fixed on the adiabat that is described by the constants $k_D$ and $k_A$ for donor and accretor. This setup is identical to that of the $q=1.3$ mass ratio setup of \citet{TohlineDD2}.

The right side of Figure \ref{f.eos} shows an ideal gas equation of state run instead. This simulation now includes shocks, which dramatically changes the dynamics of the simulation. Gas accreting from the donor (top) is strongly shocked when it hits the surface of the accretor (bottom). The major difference between the two runs is that the ideal gas equation of state builds up a hot halo of shocked material around the accretor which quickly starts to engulf the binary, essentially forming a common envelope system with different dynamics. Also note how the shocked gas in the ideal gas run is blown off the backside of both stars. This fundamentally changes the dynamics of the merger, since the expelled mass carries angular momentum outward with it. The next section will explore more details on this shocked run. Our goal is to produce quantitative comparison with the grid-based simulations of \citet{TohlineDD2} and \citet{TohlineDD3}.

\section{Case Study: A Merger with a $q=1.3$ Mass Ratio with Shocks}

\begin{figure}
\begin{center}
\includegraphics[width=0.3\textwidth]{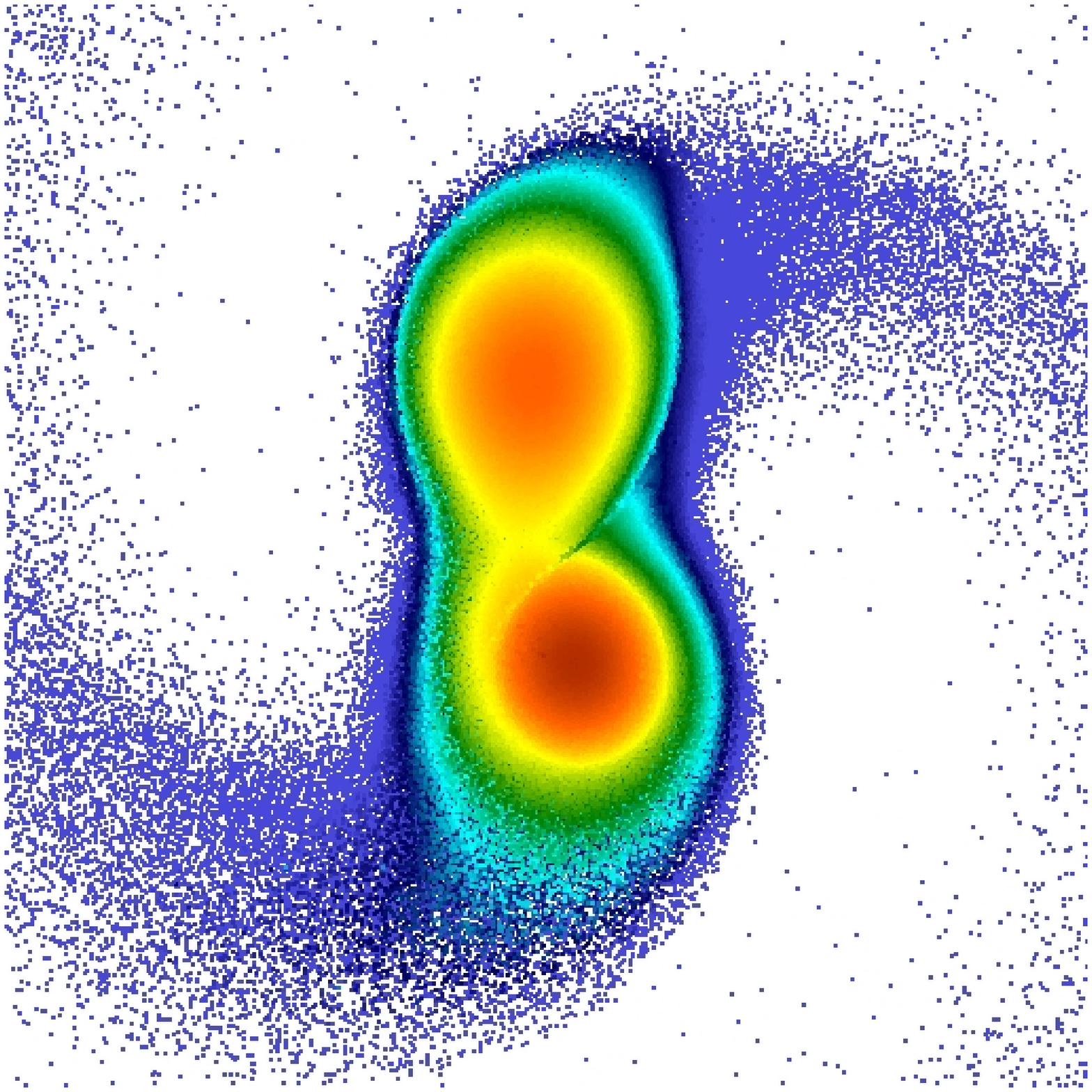}
\includegraphics[width=0.3\textwidth]{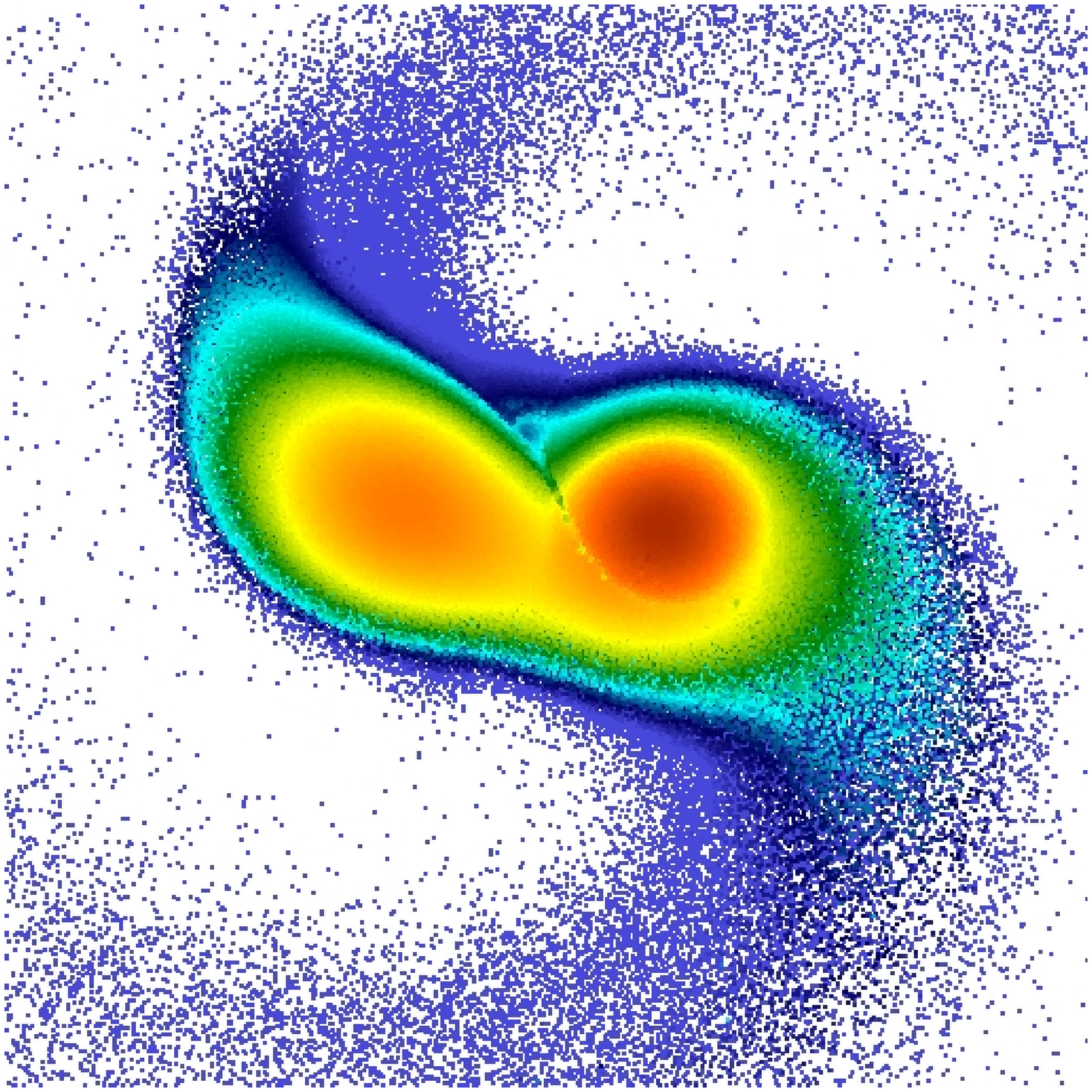}
\includegraphics[width=0.3\textwidth]{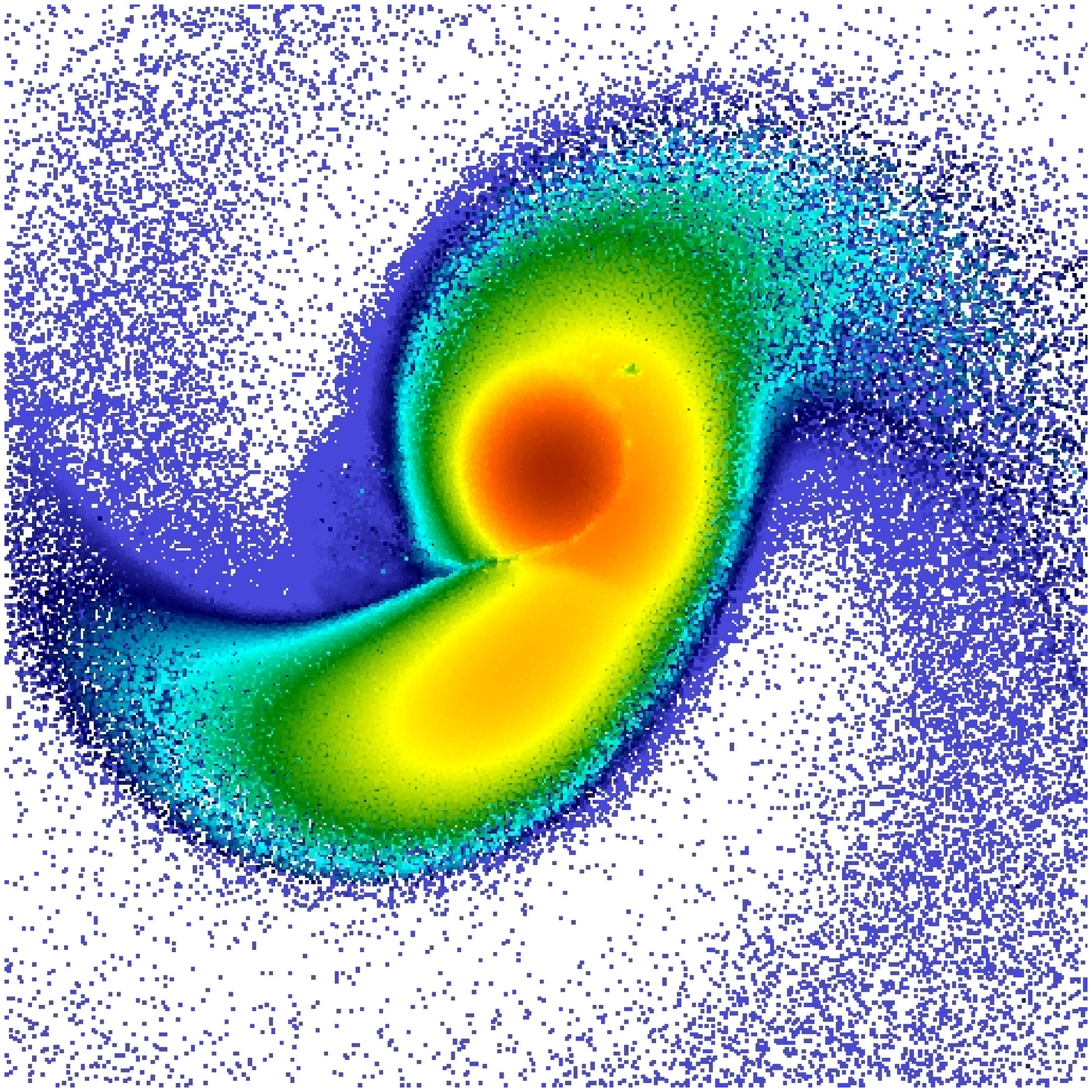}
\includegraphics[width=0.3\textwidth]{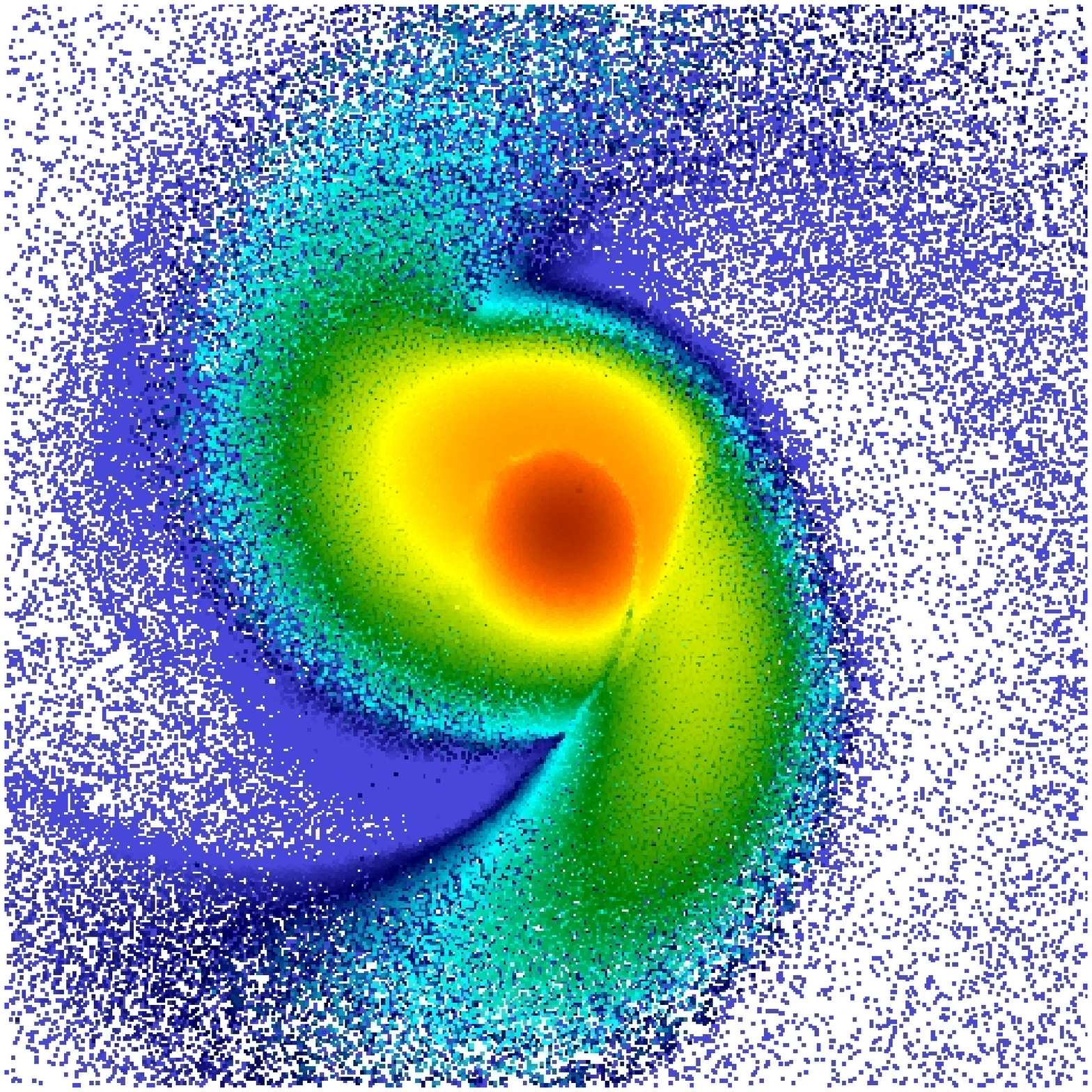}
\includegraphics[width=0.3\textwidth]{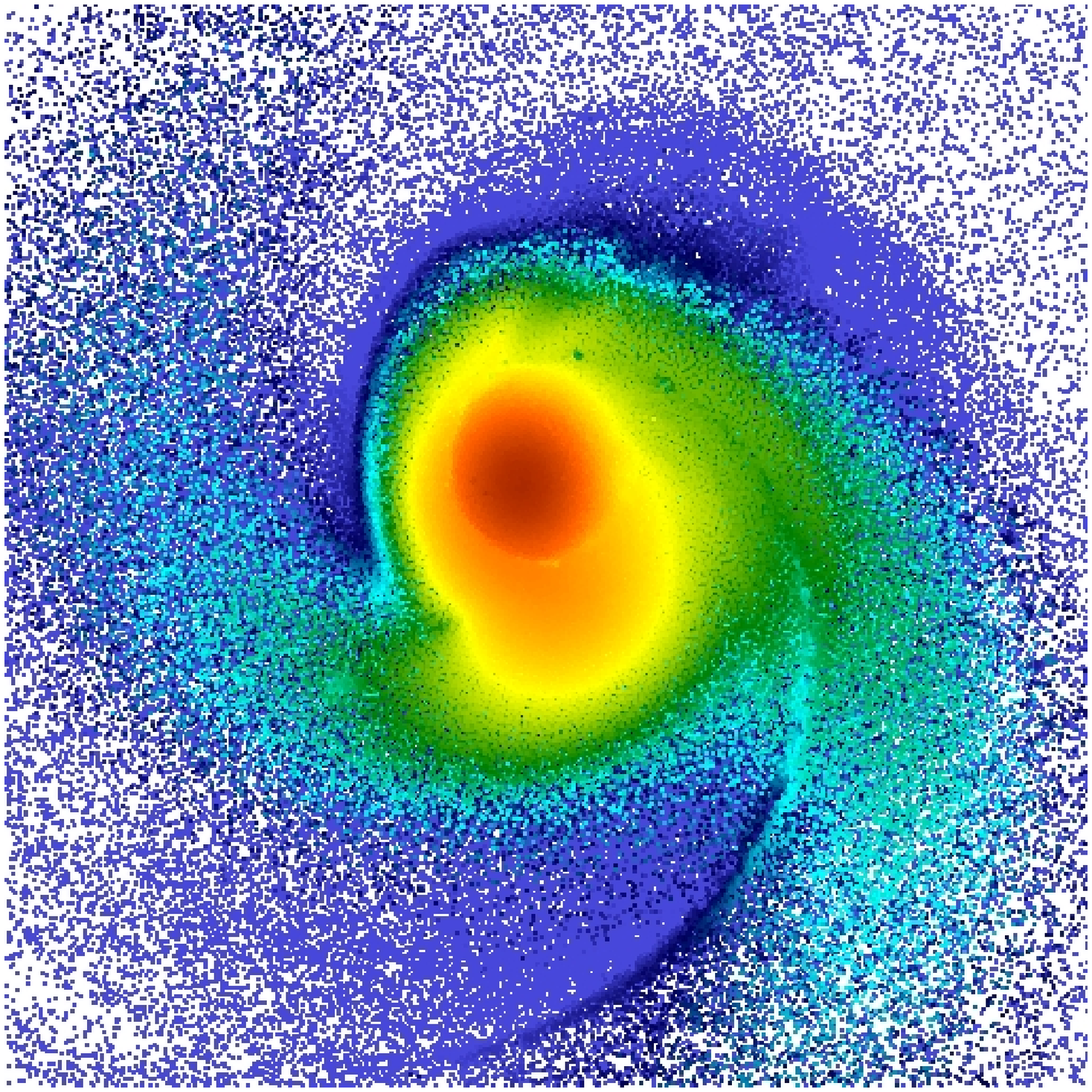}
\includegraphics[width=0.3\textwidth]{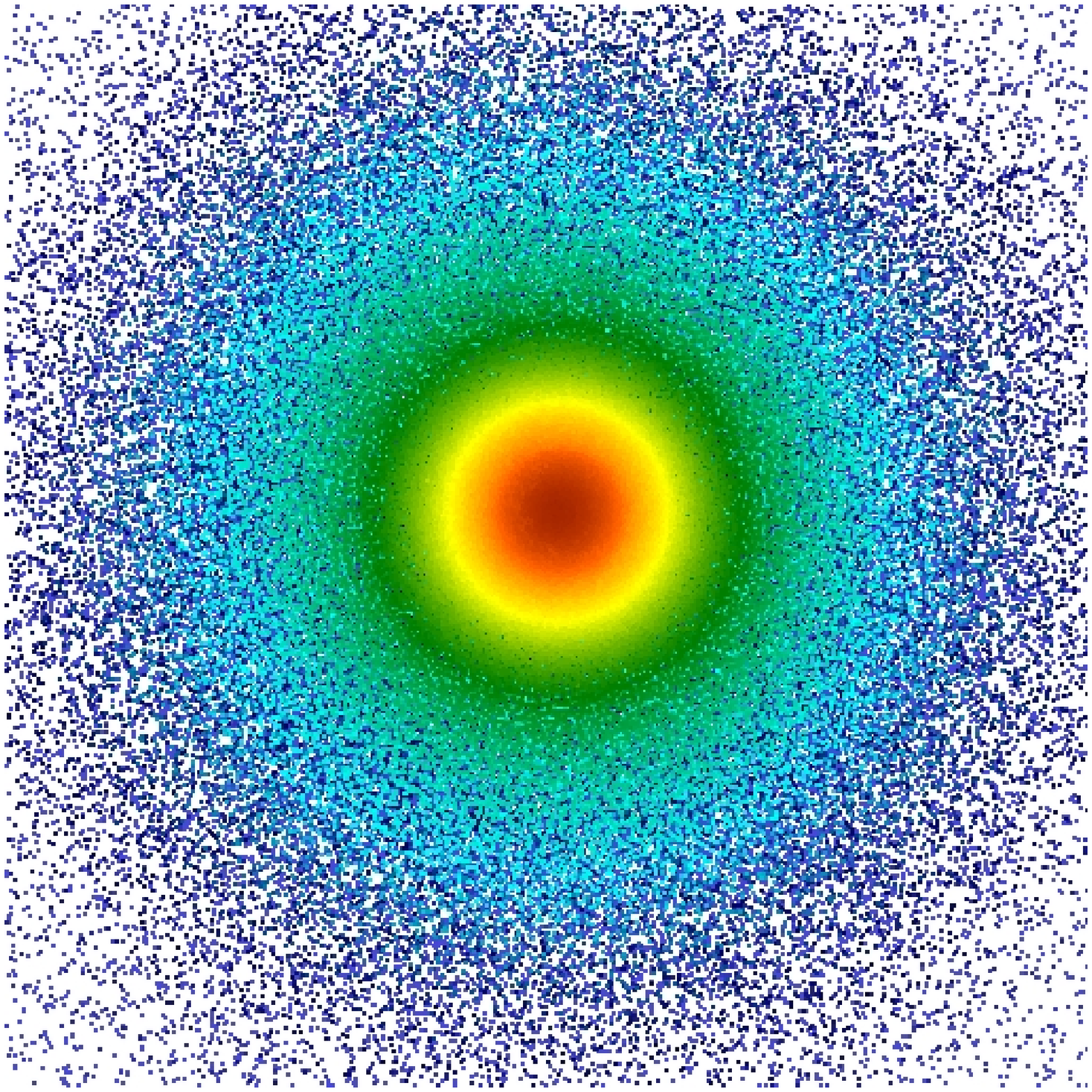}
\end{center}
\caption{SPH simulation of the $q=1.3$ merger simulation with an ideal gas equation of state. The figure show a time sequence starting at the onset of the merging process after about $9.5$ orbits. Note the importance of the strong shocks. The last panel shows the core of the relaxed merger remnant ($\sim16$ orbital periods).\label{f.Q13}}
\end{figure}

Figure \ref{f.Q13} shows the time evolution of a double-degenerate merger (rotating anti-clockwise) between two polytropes with a mass ratio of  $q=1.3$. This sequence shows the crucial 1.5 orbits where the most dynamic phase of the actual merger takes place (starting at around 9.5 orbits). This run assumes an ideal gas equation of state and includes shocks, which can be easily seen as strong density discontinuities in the snapshots. 

The last panel on the lower right shows the merger remnant at a much later stage. Note that the remnant settled into an almost spherically symmetric configuration, and is differentially rotating, with a fast rotating core and a hot envelope. At the end of our simulation, the extent of the halo is around 100 solar radii, and is still expanding. However, we do emphasize that the actual behavior may very well quantitatively differ when a realistic equation of state is employed.

\section{Conclusions and Outlook}

We are on the path to produce high-precision 3D SPH simulations of double-degenerate mergers with various mass ratios. We are carrying out a detailed verification (code comparison with grid codes, and numerical convergence study) and validation (comparison to R Coronae Borealis stars) effort. Particular emphasis is put on the initial condition setup, and the choice of the equation of state, which we both find to strongly affect the dynamics of the merger. 

Our goal is to implement more realistic equations of state to use a stellar evolution model, and to follow dynamically important nuclear reactions and their energy input in the simulations. As part of the {\it NuGrid} collaboration\footnote{\texttt{http://forum.astro.keele.ac.uk:8080/nugrid}}, we will also use the new post-processing tool \texttt{tppnp} to follow the evolution of SPH particles with a complete nuclear network, and use this output to validate our simulations with abundance measurements observed in R Coronae Borealis Stars \citep{ClaytonHdC2}. The ultimate goal of this project is to find out whether DD mergers are valid paths to type 1a supernovae and R Coronae Borealis Stars.

\bibliographystyle{apj}
\bibliography{../bibtex/allreferences.bib}

%\begin{thebibliography}{99}
%\bibitem{...} 
%....
%\end{thebibliography}

\end{document}